\documentclass[aps,prl,twocolumn,superscriptaddress]{revtex4}
\usepackage{amsbsy,amssymb,amsmath,bm}
\usepackage{graphicx,color,epsfig,rotate}
\usepackage{tabu}
\usepackage{tabularx}
\usepackage{xspace,units}
\usepackage{subfigure}
\usepackage{textcomp}		
\usepackage{epstopdf}
\usepackage{color,array}
\usepackage{array}
\usepackage[dvipsnames]{xcolor}
\usepackage{xcolor}
\usepackage{subfiles}

\makeatletter

\newcommand*{\@rowstyle}{}

\newcommand*{\rowstyle}[1]{
  \gdef\@rowstyle{#1}%
  \@rowstyle\ignorespaces%
}

\newcolumntype{=}{
  >{\gdef\@rowstyle{}}%
}

\newcolumntype{+}{
  >{\@rowstyle}%
}

\makeatother

\begin{document}

\title{Fermi-Surface Reconstruction and Dimensional Topology Change in Nd-doped CeCoIn$_5$}

\author{J. Klotz}
\affiliation{Hochfeld-Magnetlabor Dresden (HLD-EMFL), Helmholtz-Zentrum Dresden-Rossendorf, 01328 Dresden, Germany}
\affiliation{Institut f\"{u}r Festk\"{o}rper- und Materialphysik, Technische Universit\"at Dresden, 01069 Dresden, Germany}

\author{K. G\"{o}tze}
\altaffiliation{Present address: Department of Physics, University of Warwick, Coventry CV4 7AL, UK}
\affiliation{Hochfeld-Magnetlabor Dresden (HLD-EMFL), Helmholtz-Zentrum Dresden-Rossendorf, 01328 Dresden, Germany}
\affiliation{Institut f\"{u}r Festk\"{o}rper- und Materialphysik, Technische Universit\"at Dresden, 01069 Dresden, Germany}

\author{I. Sheikin}
\affiliation{Laboratoire National des Champs Magn\'{e}tiques Intenses (LNCMI-EMFL), CNRS, UGA, F-38042 Grenoble, France}

\author{T. F\"{o}rster}
\affiliation{Hochfeld-Magnetlabor Dresden (HLD-EMFL), Helmholtz-Zentrum Dresden-Rossendorf, 01328 Dresden, Germany} 

\author{D. Graf}
\affiliation{National High Magnetic Field Laboratory, Florida State University, Tallahassee, Florida 32310, USA}

\author{J.-H. Park}
\affiliation{National High Magnetic Field Laboratory, Florida State University, Tallahassee, Florida 32310, USA}

\author{E. S. Choi}
\affiliation{National High Magnetic Field Laboratory, Florida State University, Tallahassee, Florida 32310, USA}

\author{R. Hu}
\affiliation{Condensed Matter Physics and Materials Science Department, Brookhaven National Laboratory, Upton, New York 11973, USA}

\author{C. Petrovic}
\affiliation{Condensed Matter Physics and Materials Science Department, Brookhaven National Laboratory, Upton, New York 11973, USA}

\author{J. Wosnitza}
\affiliation{Hochfeld-Magnetlabor Dresden (HLD-EMFL), Helmholtz-Zentrum Dresden-Rossendorf, 01328 Dresden, Germany}
\affiliation{Institut f\"{u}r Festk\"{o}rper- und Materialphysik, Technische Universit\"at Dresden, 01069 Dresden, Germany}

\author{E. L. Green}
\email[]{e.green@hzdr.de}
\affiliation{Hochfeld-Magnetlabor Dresden (HLD-EMFL), Helmholtz-Zentrum Dresden-Rossendorf, 01328 Dresden, Germany} 

\date{\today}

\begin{abstract}

We performed low-temperature de Haas-van Alphen (dHvA) effect measurements on a Ce$_{1-x}$Nd$_x$CoIn$_5$ series, for $x =  0.02$, 0.05, 0.1, and 1, down to $T = 40$~mK using torque magnetometry in magnetic fields up to 35~T. 
Our results indicate that a Fermi-surface (FS) reconstruction occurs from a quasi-two-dimensional (2D) topology for Nd-2$\%$ to a rather three-dimensional (3D) for Nd-5$\%$, thus reducing the possibility of perfect FS nesting.
The FS evolves further with increasing Nd content with no observed divergence of the effective mass between Nd-2$\%$ and 10$\%$, consistent with the crossing of a spin density wave (SDW) type of quantum critical point (QCP). 
Our results elucidate the origin of the $Q$-phase observed at the 5$\%$ Nd-doping level \cite{Raymond14}.

\end{abstract}

\maketitle

Heavy-fermion systems have garnered attention over the past few decades due to their novel physical properties and exotic behavior, particularly when exposed to high magnetic fields \cite{Kouroudis87, Loehneysen96, Edwards97, Singh11}. 
Most notably CeCoIn$_5$, which has the highest critical temperature of all Ce-based heavy-fermion compounds ($T_c = 2.3$~K \cite{Petrovic01, Ikeda01}), exhibits a field-induced ordered phase from 10~T up to 11.4~T for $B \parallel a$ \cite{Bianchi03}, a superconducting state now referred to as the $Q$-phase. 
It was initially believed this high-field state was the realization of the long-sought-after Fulde-Ferrell Larkin-Ovchinnikov (FFLO) phase \cite{Bianchi03, Fulde, Larkin, Radovan03}, however, further measurements revealed the presence of antiferromagnetic (AFM) order \cite{Mitrovic06, Young07, Kenzelmann08}. 
Many questions, however, still remain about the interactions appearing in this high-field superconducting state.

By substituting a portion of the Ce atoms in CeCoIn$_5$ with Nd, an atom where the 4$f$ electrons do not hybridize with the conduction electrons, one can introduce SDW and, at higher Nd contents, AFM order \cite{Hu08, Raymond14}.
It was recently reported that by replacing 5$\%$ of the Ce atoms with Nd the high-field $Q$-phase could be stabilized at zero applied magnetic field, inciting intense research to explore the mechanism behind such unique properties \cite{Raymond14, Mazzone17, Mazzone_PRL_17, Mazzone18}. 
An enhancement of the nesting has been proposed \cite{Raymond14}, however, experimental proof has been insufficient, thus providing our motivation to explore how Nd alters the FS.  
Additionally, it has previously been suggested that a QCP may exist in these materials \cite{Mazzone17}, which may also be induced by increasing Nd doping \cite{Hu08}.


In this paper, we provide evidence for a drastic FS reconstruction between 2$\%$ and 5$\%$ Nd substitution. 
The reconstruction manifests itself in a sharp deviation of the $\alpha$ orbit from the expected $1/\cos\theta$ dependence normally associated with a quasi-2D cylindrical FS. 
This is at odds with assumptions made in current theoretical models \cite{Martiny15}.
Furthermore, this distorted FS may be described as more 3D in nature and any effects from possible FS nesting suggested in prior works \cite{Raymond14} may, therefore, be significantly reduced.
The FSs continue to evolve for 5\%, 10\%, and 100\% Nd substitution and are consistent with fully localized $4f$ electrons only at 100\% substitution.
The effective mass associated with the quasi-2D cylindrical band, previously found to play a key role in superconductivity \cite{Allan13, Goetze15}, remains constant between 0 and 10$\%$ Nd substitution levels.
Both the progressive change of the FSs with increasing substitution level and non-divergent effective carrier masses indicate the SDW-type nature of the QCP \cite{Si10, Gegenwart08}, consistent with thermodynamic measurements \cite{Hu08}.

\begin{figure}
\centering
\includegraphics[width=0.95\columnwidth]{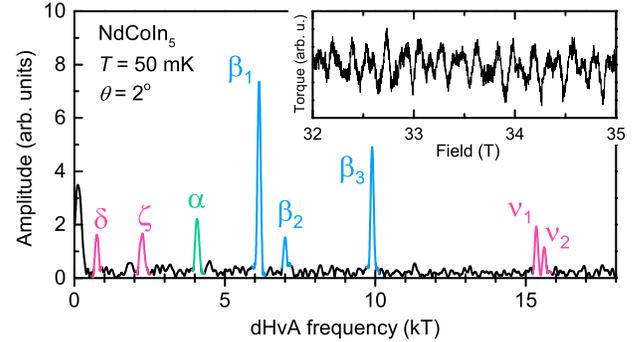}
\caption{Fourier transform of the quantum oscillations between 21-35~T measured on NdCoIn$_5$ at an angle 
$\theta$ = 2$^\circ$ from [001] to [110]. Inset shows the torque signal at high magnetic fields after background
subtraction.}
\label{fig:Fig1}
\end{figure}

\noindent%
\begin{figure*}[ht!]  
\begin{minipage}{0.5\textwidth}
\centering
\includegraphics[width=0.8\textwidth]{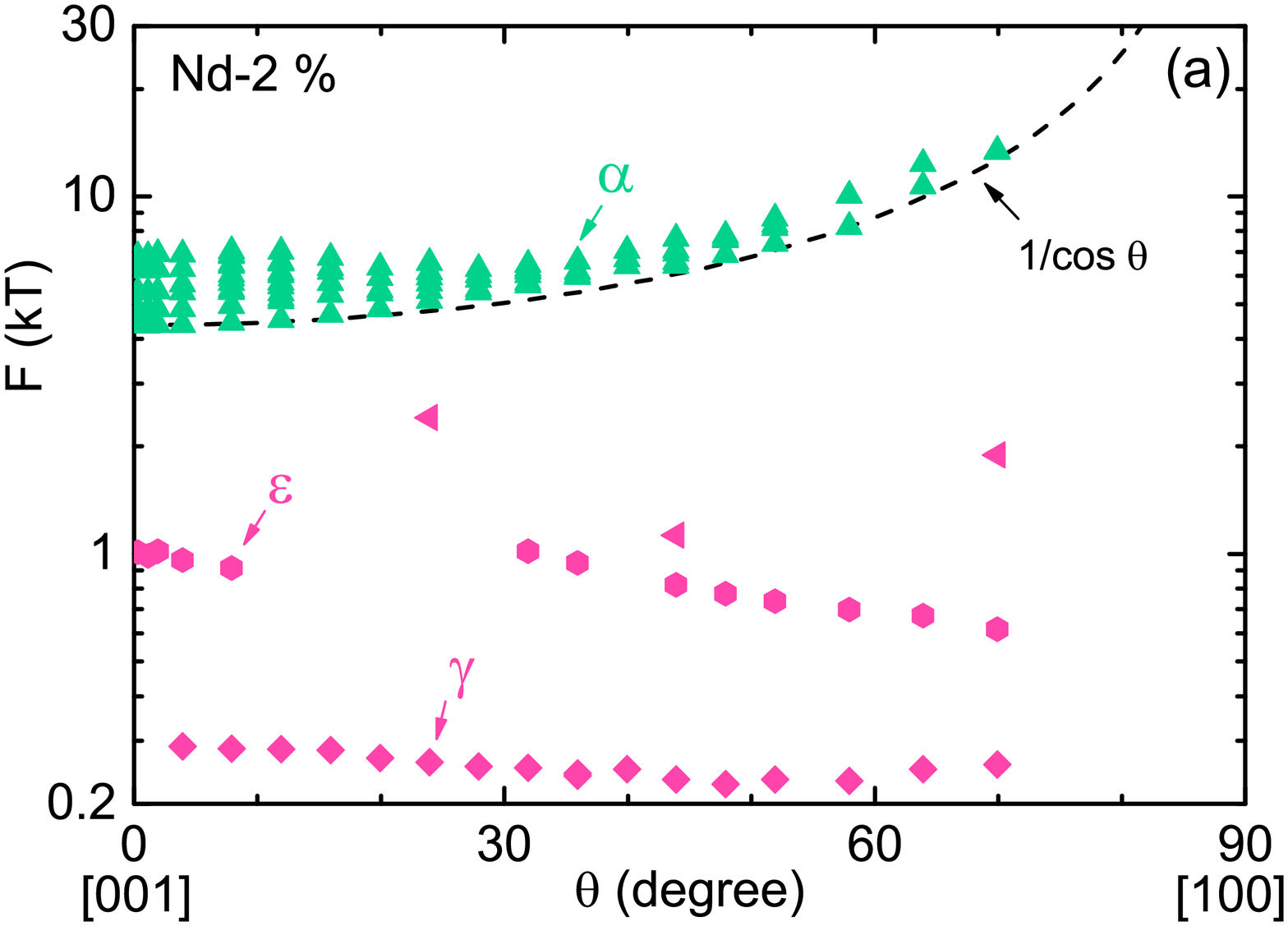}
\end{minipage}%
\noindent%
\begin{minipage}{0.5\textwidth}
\centering
\includegraphics[width=0.8\textwidth]{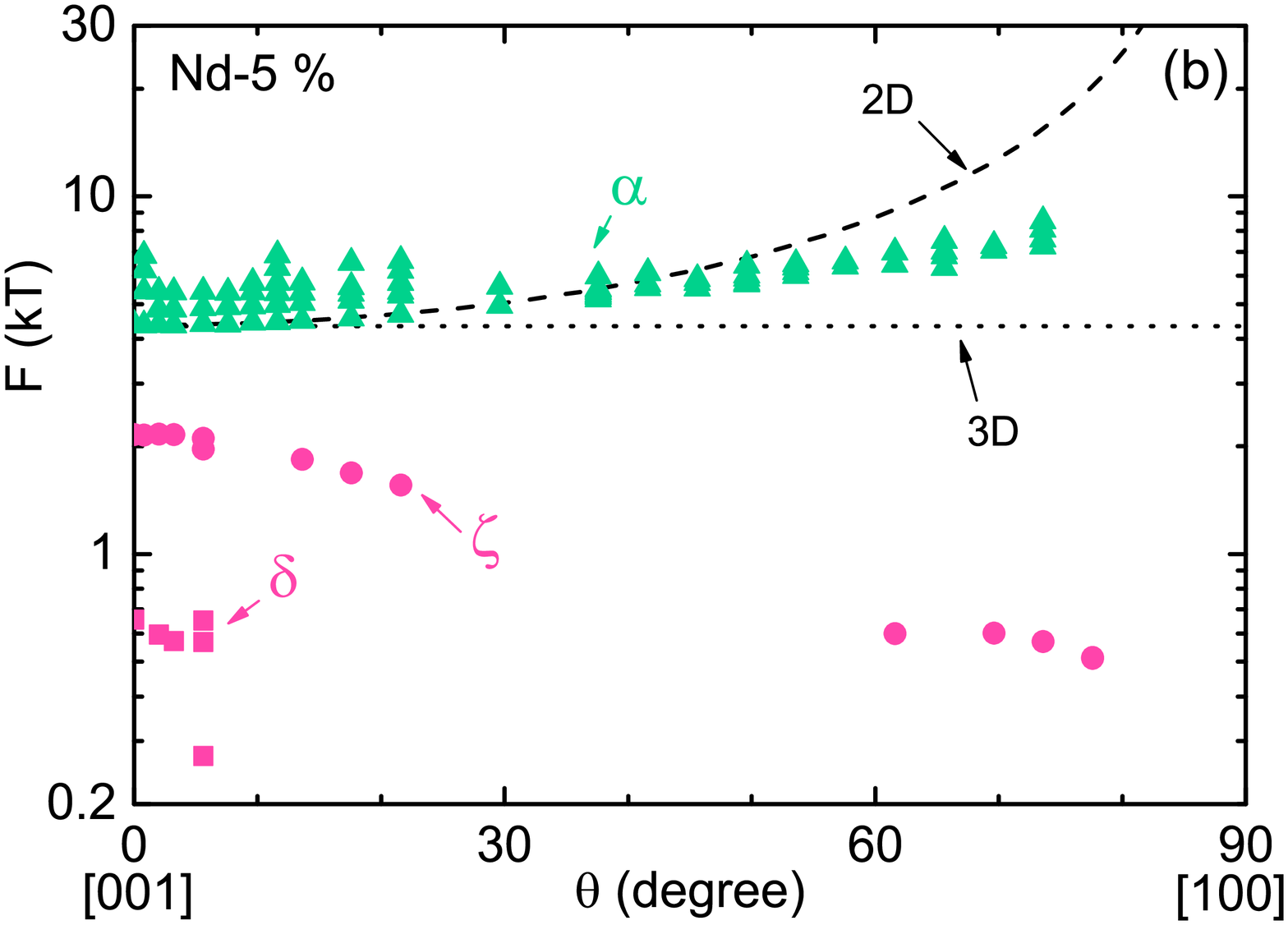}
\end{minipage}%
\vspace{0.2cm}
\noindent%
\begin{minipage}{0.5\textwidth}
\centering
\includegraphics[width=0.8\textwidth]{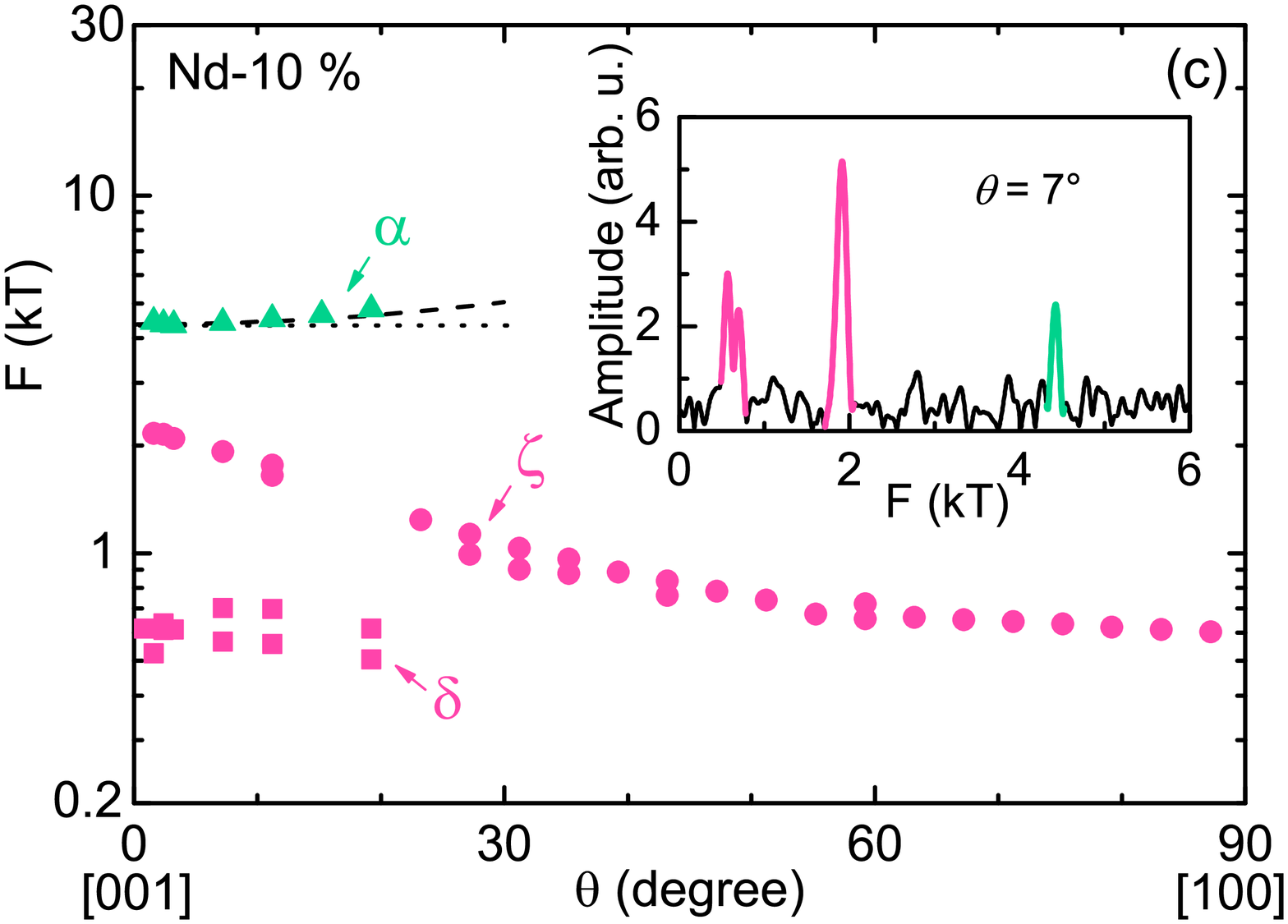}
\end{minipage}%
\noindent%
\begin{minipage}{0.5\textwidth}
\centering
\includegraphics[width=0.8\textwidth]{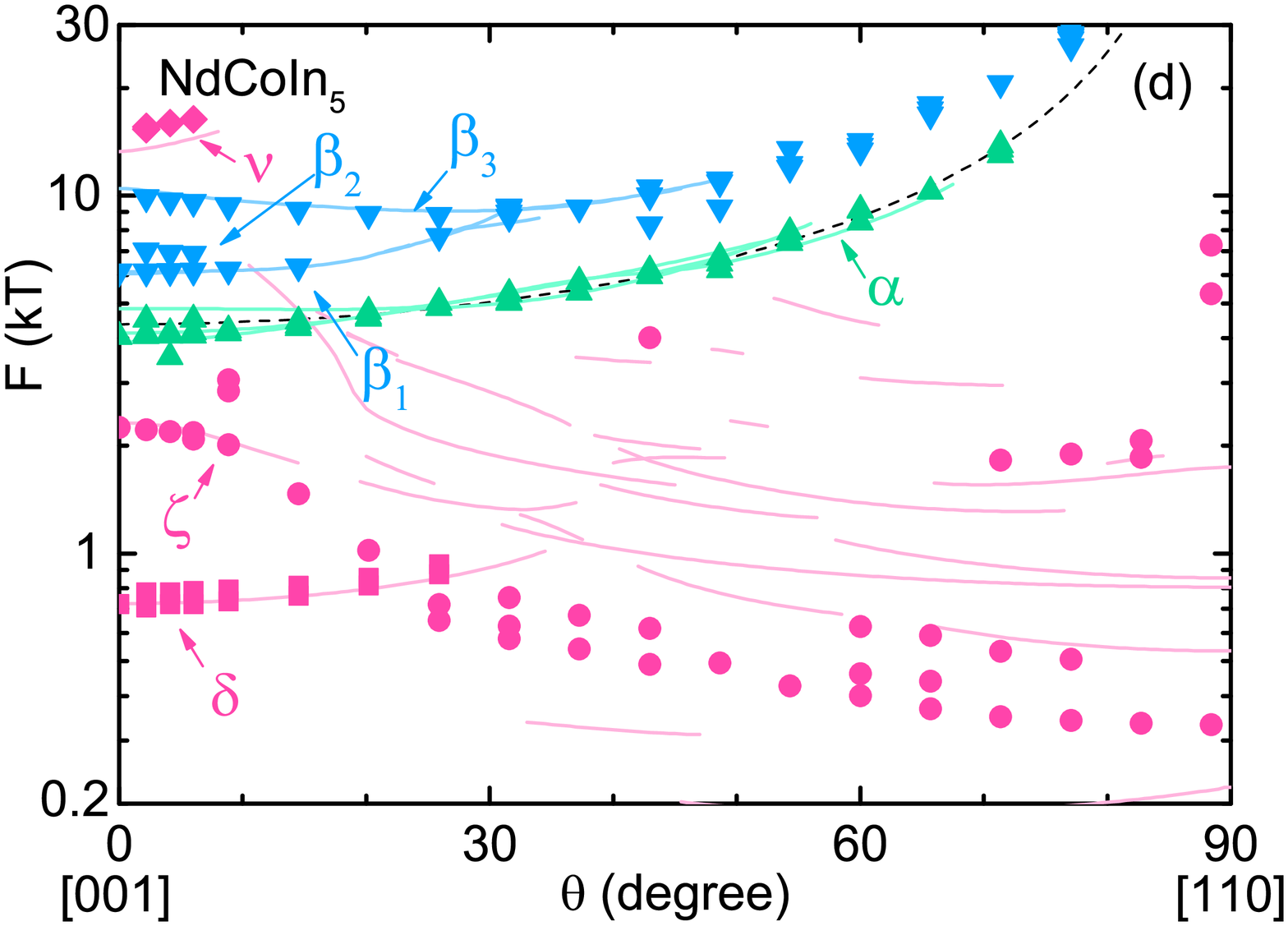}
\end{minipage}%
\caption{Angular dependence of the dHvA frequencies for (a) Nd-2$\%$, (b) 5$\%$, (c) 10$\%$, and (d) 100 $\%$. The latter includes the results of the band-structure calculations, shown as lines. 
The labeling of the frequency branches has been defined so as to be consistent with prior works \cite{Hall01, Polyakov12}. 
Dashed lines denote 1/cos $\theta$ dependence arising from a 2D cylindrical FS, dotted from a 3D sphere. 
Inset in (c) depicts the FFT for $\theta$ = 7.2$^\circ$.}
\label{fig:Fig2}
\end{figure*}

High-quality single crystals of Ce$_{1-x}$Nd$_x$CoIn$_5$ with $x$ = 0.02, 0.05, 0.1, and 1 (herein referred to as Nd-2$\%$, 5$\%$, 10$\%$, and 100$\%$) were grown and characterized as described elsewhere \cite{Hu08}.
Torque magnetometry was performed at the Laboratoire National des Champs Magn\'{e}tiques Intenses (LNCMI) in Grenoble, France, and the National High Magnetic Field Laboratory (NHMFL) in Tallahassee, USA, using a capacitive beam cantilever mounted on a rotator and placed in a top-loading $^3$He/$^4$He dilution refrigerator.
To accurately probe the lighter effective masses in NdCoIn$_5$, a $^3$He cryostat was utilized. 
Band-structure calculations for NdCoIn$_5$ were performed using the full-potential local-orbital (FPLO) minimum-basis code (version 9.01-35) \cite{koepernik_FPLO_1999} with a scalar-relativistic setting and the local density approximation of Ref.~\cite{perdew_wang_1992}. 
Lattice parameters were taken from Ref.~\cite{hudis_NdCoIn_growth_2006}.

\begin{figure}
\centering
\includegraphics[width=0.9\columnwidth]{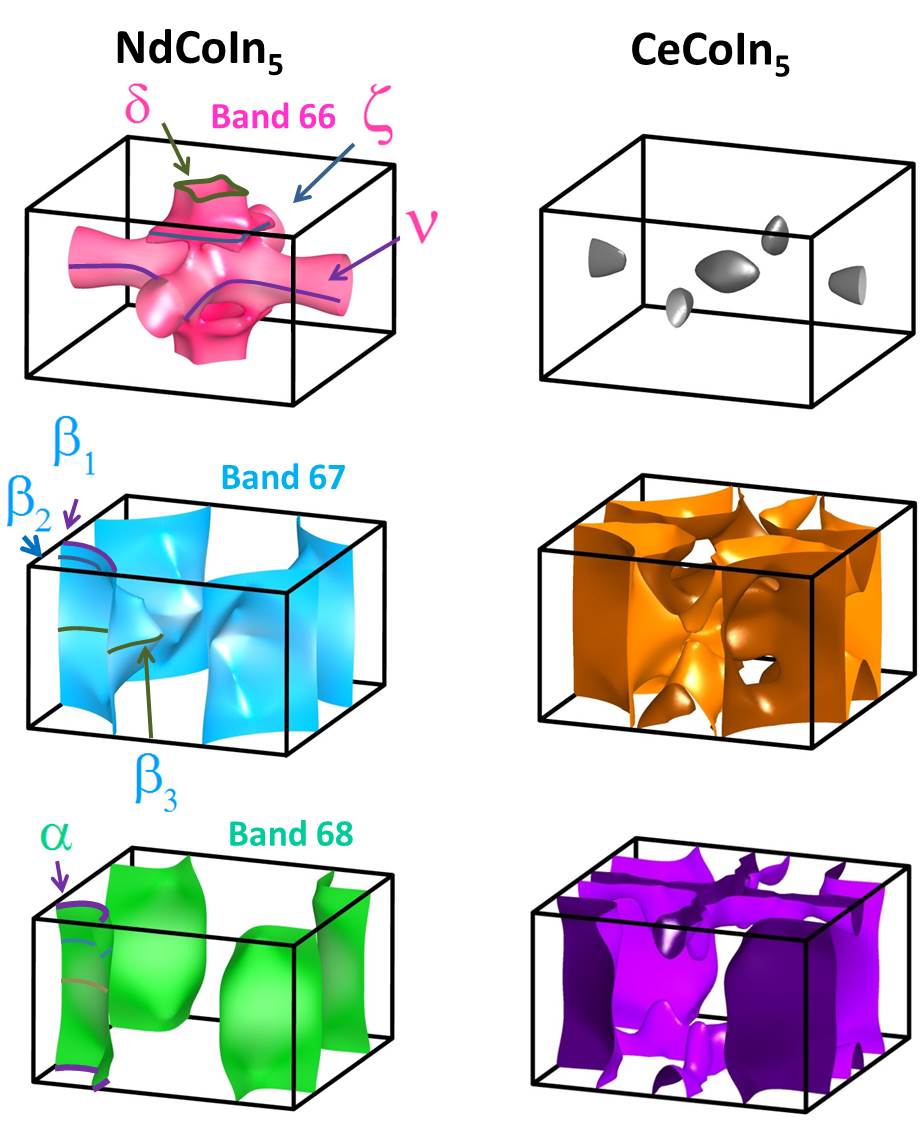}
\caption{
Calculated FSs for NdCoIn$_5$ (left) and CeCoIn$_5$ (right). The black frames depict the first Brillouin zone. 
The solid lines on the FSs aid in the visualization of some extremal orbits for the magnetic field aligned along the $c$ axis.
For NdCoIn$_5$ (left) 66 is a hole band while 67 and 68 are electron bands.}
\label{fig:Fig3}
\end{figure}

dHvA oscillations were detected using field-dependent torque magnetometry (21-35~T) for Nd-2$\%$, 5$\%$, 10$\%$, and 100$\%$, but could not be resolved for mid-range substitution levels (Nd-20$\%$ and 60$\%$).
Background magnetization was subtracted using a higher-order polynomial fit resulting in oscillations as shown for NdCoIn$_5$ in the inset of Fig.~\ref{fig:Fig1}. 
A Fourier transform revealed several prominent frequencies well above the noise level (see Fig.~\ref{fig:Fig1}). 
Angular dependence for the dHvA frequencies was explored, rotating from $B \parallel c$ ([001]) to $B \perp c$ ([100]), for Nd-2$\%$, 5$\%$, and 10$\%$ as shown in Figs.~\ref{fig:Fig2}(a)-(c). 
Nd-$100\%$ was rotated towards [110] [see Fig.~\ref{fig:Fig2}(d)].
In pure CeCoIn$_5$, it has been shown that there is little difference in the dHvA frequencies between [100] and [110] due to the mostly isotropic FSs in the $a$-$b$ plane \cite{Settai01, Hall01}. 
The same is assumed to be true for the isostructural compound NdCoIn$_5$, backed by our calculations.

\begin{figure}[h!]
\includegraphics[width=0.9\columnwidth]{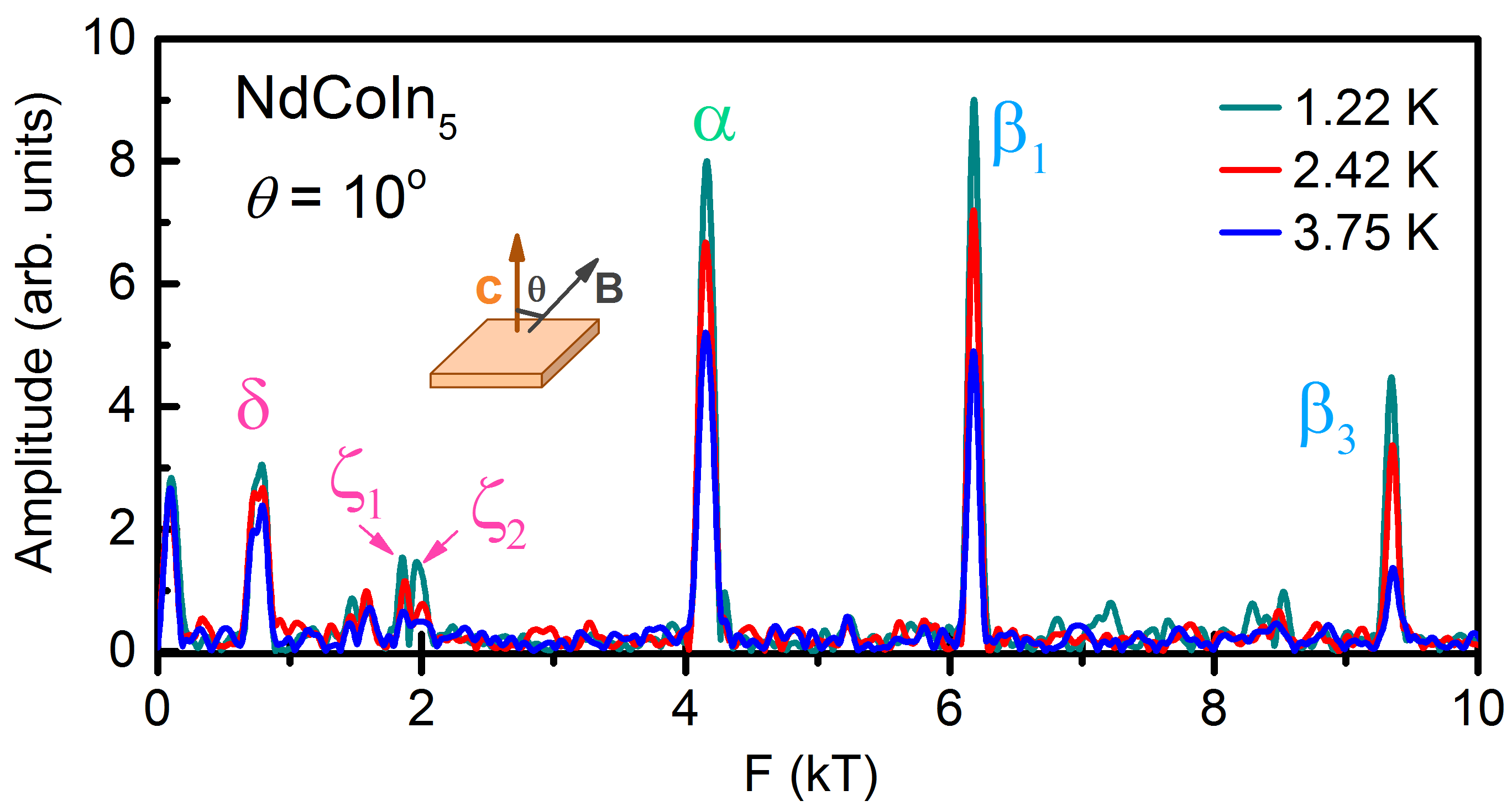}
\caption{Example of the temperature dependence of the observed frequencies in NdCoIn$_5$ measured between 21-35~T
at an angle of $\theta = 10^\circ$ from [001] to [110].}
\label{fig:Fig5}
\end{figure}

Numerous frequencies could be resolved for Nd-2$\%$, as shown in Fig.~\ref{fig:Fig2}(a), many of which correspond to those found in CeCoIn$_5$ \cite{Hall01, Shishido03}. 
The frequency labeled $\beta$ by Settai $et$ $al.$~\cite{Settai01} could not be observed, possibly due to increased impurity scattering, which affects heavy-mass orbits stronger. 
Instead, we observe frequencies of 6.25 and 6.92~kT at $\theta=3^\circ$, which were not reported for pure CeCoIn$_5$ \cite{Settai01, Hall01, Shishido03, Polyakov12}. 
Since the frequencies between 4 and 8~kT, labeled $\alpha$, follow roughly a 1/cos\ $\theta$ dependence, where $\theta$ is the angle between the applied magnetic field and the $c$ direction (see drawing in Fig.~\ref{fig:Fig5}), it is plausible that these frequencies arise from a strongly corrugated 2D cylindrical FS. 

A drastic FS reconstruction occurs for 5$\%$ Nd doping [Fig.~\ref{fig:Fig2}(b)].
The frequencies $\epsilon$ and $\gamma$ in Nd-2$\%$ disappear and are replaced by $\delta$ and $\zeta$ in Nd-5$\%$.
Frequencies associated with the $\alpha$ orbits noticeably deviate from the expected 1/cos $\theta$ dependence seen in the 2$\%$ Nd-doping level indicating the 2D cylindrical FS has now warped into more of a quasi-3D structure. 
This is of particular interest because recent theoretical work exploring the reasoning behind the appearance of the $Q$-phase at zero applied magnetic field is based on the assumption that doping has a negligible effect on the FS \cite{Martiny15}. 
Furthermore, it has been proposed that the $Q$ vector appears at 0~T due to an enhanced FS nesting \cite{Raymond14}. 
However, since the cylindrical FS is developing a more 3D structure, this scenario seems unlikely.  

As the substitution level increases to 10$\%$, the frequencies associated with the $\delta$ and $\zeta$ continue to evolve [Fig.~\ref{fig:Fig2}(c)]. 
Few frequencies associated with the $\alpha$ band could be seen, all agreeing with frequencies found for Nd-2$\%$ and Nd-5$\%$. 
Due to increased scattering provoked by doping-induced disorder, quantum oscillations could not be detected for further substitution levels (20$\%$ and 60$\%$), but numerous frequencies appear for NdCoIn$_5$ as seen in Fig.~\ref{fig:Fig2}(d). 
Furthermore, a well-defined 2D cylindrical FS, labeled $\alpha$ as before, reemerges as evidenced by the 1/cos $\theta$ behavior, but with a different FS cross section.
In addition, a second corrugated 2D cylindrical FS labeled $\beta$ is identified by a $1/\cos\theta$ dependence.

Band-structure calculations for NdCoIn$_5$, performed assuming localized, non-hybridizing 4$f$ electrons, are in good agreement with experimental data [see Fig.~\ref{fig:Fig2}d]. 
The topology of the FSs for NdCoIn$_5$ along with their corresponding frequencies are depicted in Figs.~\ref{fig:Fig2}(d) and \ref{fig:Fig3}, herein referred to as band 66, band 67, and band 68 represented in the graphs as pink, blue, and green, respectively.
A comparison between the FSs of CeCoIn$_5$ and NdCoIn$_5$ is shown in Fig.~\ref{fig:Fig3}.
As a note, the FS could only reasonably be calculated for NdCoIn$_5$ and CeCoIn$_5$, not for the substitution series because of the increased complexity doping introduces to the calculation. 

Effective masses were extracted by fitting the oscillation amplitudes, shown in Fig.~\ref{fig:Fig5}, to the Lifshitz-Kosevich formula \cite{Shoenberg, Suppl_Mat}. 
In order to observe a sufficiently strong torque signal for as many frequency branches as possible, effective masses were measured at angles between 4 and 10$^\circ$ from the $c$ axis for the different substitution levels. 
Effective masses for the observed frequencies in NdCoIn$_5$ are shown in Table \ref{tab:NdCoIn_masses} and are slightly larger than the calculated masses reflecting modest mass enhancements quantified by $\lambda=m^\ast / m_b -1$ \cite{note_1}.
Effective masses for all other substitution levels are shown in Table \ref{tab:Ndx_masses}. 
It is evident that increasing the substitution level has little effect on the effective masses of existing orbits up to Nd-10$\%$, while newly appearing orbits have a greatly reduced mass.

The effective mass for $\alpha_3$, the orbit associated with the 2D band at low doping level and the only branch that can be tracked  up to Nd-10$\%$, remains relatively unchanged with increasing substitution levels despite $T_c$ rapidly decreasing 
($T_c$ = 2.1, 2.0, and 1.8 K for Nd-2\%, 5\%, and 10\%, respectively) \cite{Hu08}.  
This behavior near QCPs was observed in the CeRh$_{1-x}$Co$_x$In$_5$ \cite{Goh08} and CeCo(In$_{1-x}$Cd$_x$)$_5$ \cite{Capan10} series, but differs from the Ce$_x$La$_{1-x}$CoIn$_5$ \cite{Harrison04} series and the pressure-induced QCP in CeRhIn$_5$ \cite{Shishido05}.
Consequently, our data indicate the presence of a SDW-type QCP \cite{Si10, Gegenwart08}, consistent with the experimental phase diagram \cite{Hu08}.
The cylindrical FSs are generally assumed to be the heavy bands that possess the largest energy gap \cite{Allan13} and are theoretically considered to experience the interacting pairing potential relevant for superconductivity \cite{Dyke14}.
Therefore, since the band remains relatively heavy, most probably due to electronic correlations, Nd is likely altering the pairing potential.
This is in agreement with the reduced jump in the specific heat upon entering the superconducting state \cite{Hu08}.

\begin{table}[ht]
   \centering
    \begin{tabular*}{\columnwidth}{@{\extracolsep{\stretch{1}}}*{7}{c | c c | c c | c}@{}}
       \hline
       \hline
                      & \multicolumn{2}{c|}{Experiment}  & \multicolumn{2}{c|}{Calculation} &	$\lambda$ \\
                     branch & $F$ [kT] & $m^*$ [$m_e$]	&  $F$ [kT]& $m_b$ [$m_e$] & $m^*/m_b-1$ \\				 
			\hline
			  \rule{0pt}{1\normalbaselineskip}      $\delta$	&	 0.79		&	 0.53(8)	&	 0.74		&	 0.29	 & 0.82	\\
			      $\zeta{_1}$	&	 1.87		&	 1.0(2)		&						&	 &       				\\
			       $\zeta{_2}$	&	 1.98		&	 1.2(2)		&						&	  &      				\\
			   $\alpha$	&	 4.15	        &	0.83(3)		&	4.12	       & 	0.73	&  0.13	\\			  
			          $\beta{_1}$	&	 6.17		&	 1.06(3)	&	 6.18		&	 0.77	&  0.37	\\
			          $\beta{_3}$	&	 9.36		&	 1.3(2)		&	 9.42		&	 0.86	& 0.55	\\
       \hline
       \hline
     \end{tabular*}
     \caption{Comparison of effective masses measured at an angle of 10$^\circ$ from [001] to [110] of NdCoIn$_5$ with calculations. }
\label{tab:NdCoIn_masses}
\end{table}

\begin{table*}[t]
	\centering
		\begin{tabular*}{\textwidth}{@{\extracolsep{\stretch{1}}}*{9}{c | c c | c c | c c | c c}@{}}
			\hline
			\hline
				\rule{0pt}{1\normalbaselineskip}& \multicolumn{2}{c|}{CeCoIn$_5$\cite{Polyakov12}} &  \multicolumn{2}{c|}{Nd-2$\%$} & \multicolumn{2}{c|}{Nd-5$\%$}  & \multicolumn{2}{c}{Nd-10$\%$} \\
				\rule{0pt}{1\normalbaselineskip}& \multicolumn{2}{c|}{$\theta=0^\circ$}  &  \multicolumn{2}{c|}{$\theta=3^\circ$} & \multicolumn{2}{c|}{$\theta=2^\circ$}  & \multicolumn{2}{c}{$\theta=7^\circ$} \\
			\hline
				\rule{0pt}{1\normalbaselineskip} 		branch	& $F$ [kT] 	& \multicolumn{1}{c|}{$m^*$ $[m_e]$} &  $F$ [kT] 	& \multicolumn{1}{c|}{$m^*$ $[m_e]$}	& $F$ [kT] & \multicolumn{1}{c|}{$m^*$ $[m_e]$}	& $F$ [kT] & \multicolumn{1}{c}{$m^*$ $[m_e]$}  \\
			\hline
				\rule{0pt}{1\normalbaselineskip}		$\gamma$& &  &	0.29	&	 6.4(6)		&		&	 		&		&	 	\\
										$\epsilon$& & &	0.99	&	 20.3(15)	&		&			&		&	 	\\
										$\delta{_1}$& &&		&	 		&	0.59	&	$<1$ 		&	0.55	&	$<1$ 	\\
										$\delta{_2}$& & &		&			&		&			&	0.68	&	$<1$	\\
										$\zeta$	& & &		&			&	2.16	&	$<1$ 		&	1.9	&	$<1$	\\
										$\alpha{_1}$ & 5.46 & 14.5 &	5.44	&	10.3(5)		&	5.42	&	10.0(5)		&		&		\\
										$\alpha{_2}$& 4.87 & 11.4 &	4.85	&	11.0(5)		&	4.84	&	10.5(8)		&		&		\\
										$\alpha{_3}$ & 4.37 & 9.3  & 	4.39	&	 6.7(2)		&	4.38	&	6.8(2) 		&	4.41	&	7(1)	\\
										$\alpha{_4}$& & 		 & 6.25 &	12.0(10)	&		&			&		&		\\
										$\alpha{_5}$& & 	 & 6.92	&	18.5(7)		&		&			&		&		\\
			\hline
			\hline
		\end{tabular*}
		\caption{Effective masses for the various frequencies depicted in Figs. \ref{fig:Fig3}(a)-(c). Frequencies and effective masses associated with CeCoIn$_5$ were taken from Ref.~\cite{Polyakov12}.
		$m^*$ associated with $\alpha$ remains relatively stable with increased Nd-doping.}
\label{tab:Ndx_masses}
\end{table*}


In conclusion, the FS of Ce$_{1-x}$Nd$_x$CoIn$_5$ changes across the QCP associated with the appearance of long-range magnetic order at Nd-5\% and continues to evolve with increasing $x$, consistent with a SDW-type QCP~\cite{Si10, Gegenwart08}.
Furthermore, between Nd-2$\%$ and Nd-5$\%$, the FS moves away from two-dimensionality which is at odds with an enhanced Fermi-nesting scenario given as an explanation for the $Q$-phase seen at zero applied magnetic field \cite{Raymond14}. 
Since effective masses are unaffected by altering Nd content up to 10\% despite a decrease in $T_c$, it is plausible to conclude that Nd alters the electronic pairing potential. 
Our results provide the first information on the evolution of the FS topology under Nd perturbation and its influence on the emerging 
$Q$-phase in Nd-5$\%$.



We acknowledge the support of HLD at HZDR and LNCMI at CNRS, members of the European Magnetic Field Laboratory (EMFL) and ANR-DFG grant Fermi-NESt. 
A portion of this work was performed at the National High Magnetic Field Laboratory, which is supported by National Science Foundation Cooperative Agreement No.\ DMR-1157490 and the State of Florida.
Work at Brookhaven is supported by the US DOE under Contract No. DE-SC0012704. 
C.\,P.\ acknowledges support by the Alexander von Humboldt Foundation. 
K.\,G.\ acknowledges support from the DFG within GRK 1621.
We would like to thank A.\ Carrington, R.\,R.\ Urbano, G.\ Zwicknagl, C.\ Putzke, T.\ Murphy, and H.\ Rosner for fruitful discussions.



\begin{thebibliography}{99}
\bibitem{note_1}
FPLO typically results in lower effective masses than are observed experimentally, as is the case here, because it does not take many-body interactions into account.
\end{thebibliography}
\begin{thebibliography}{99}
\bibitem{note_2}
Sommerfeld coefficients for doping levels 10$\%$ and below were estimated using the electronic specific-heat values, $C_{el}/T$, at $T = 2.5$~K.
\end{thebibliography}

\newpage
\documentclass[aps,prl,twocolumn,superscriptaddress]{revtex4}

\setcounter{figure}{0}
\renewcommand{\thefigure}{S\arabic{figure}}

\makeatletter

\makeatother

\maketitle

\section{Supplemental Material}
\subsection{for "`Fermi-Surface Reconstruction and Dimensional Topology Change in Nd-doped CeCoIn$_5$"' by J. Klotz, K. G\"{o}tze$^*$, I. Sheikin, T. F\"{o}rster, D. Graf, J.-H. Park, E. S. Choi, R. Hu, C. Petrovic, J. Wosnitza, and E. L. Green$^{\dagger}$}

\section{Calculated band structure and density of states}

The resulting energy bands from band-structure calculations using the \textsc{fplo} code for NdCoIn$_5$ are shown in Fig.~\ref{fig:Fig4}. 
Density of states are also shown with the atomic contributions, revealing the greatest contribution at the Fermi energy level arises from the $d$ electrons associated with the Nd and Co atoms and $p$ electrons from the In atoms. 

\begin{figure}[h!]
\includegraphics[width=0.99\columnwidth]{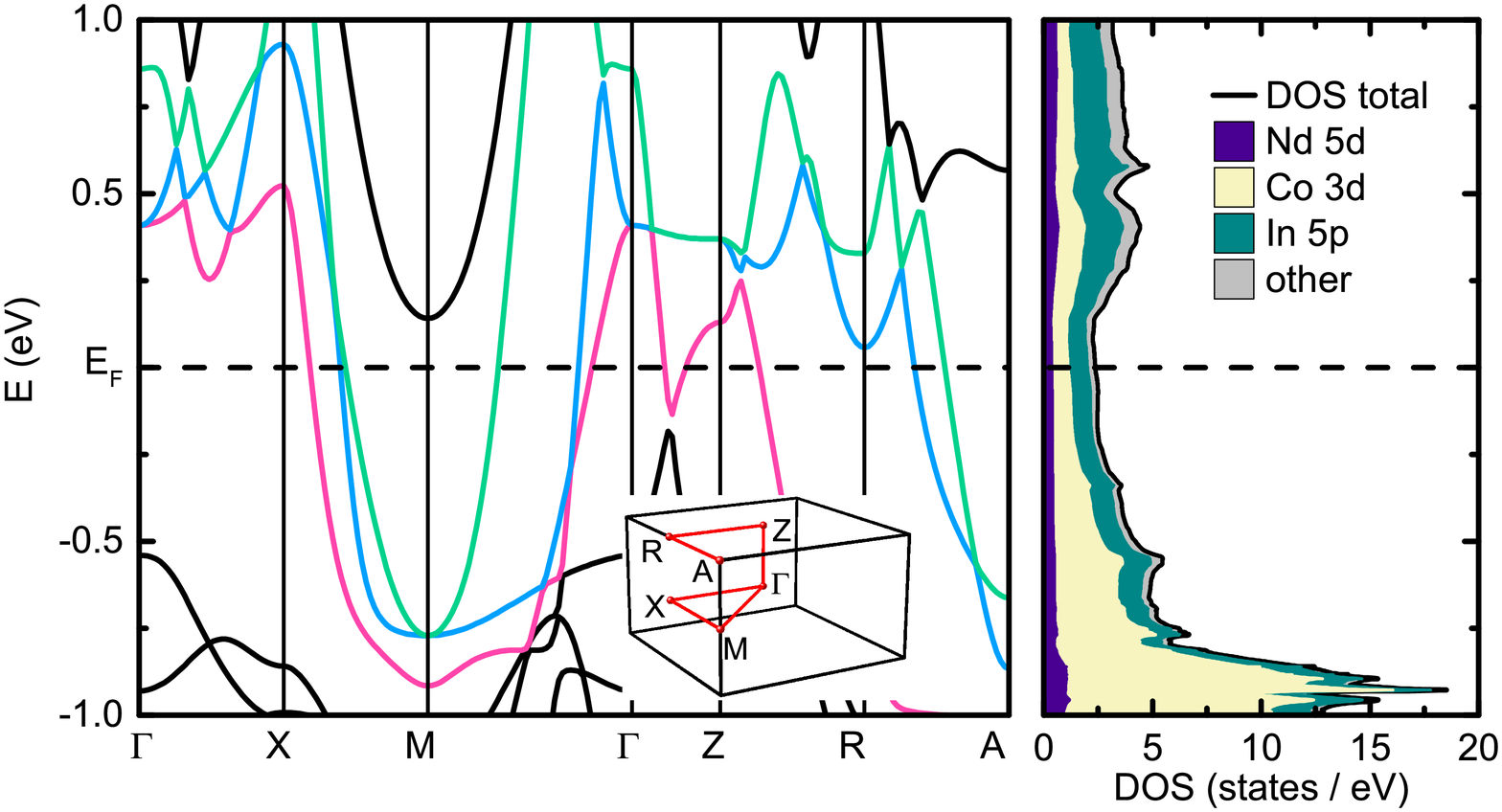}
\caption{(left) Energy bands from band-structure calculations for NdCoIn$_5$ along some high-symmetry axes of the Brillouin zone (inset). 
Highlighted dispersion relations crossing the Fermi energy are color-coordinated with their associated FSs shown in Fig.~3 of the main text.
(right) Color-coded density-of-state contributions per formula unit of the individual atomic orbitals near the Fermi energy (dashed black line).}
\label{fig:Fig4}
\end{figure}

\section{Effective-mass determination}

Effective masses were determined from the temperature dependence of the dHvA oscillation amplitudes as shown in Fig.~\ref{fig:Fig5b}.
According to the Lifshitz-Kosevich formula, this dependence is proportional to $x/\sinh x$, where $x = \alpha T m^\ast/B$ and $\alpha = 14.69$~T/K \cite{Shoenberg}.
Here $m^\ast$ represents the effective mass given in multiples of the bare electron mass $m_e$.
$T$ and $B$ represent temperature and magnetic field, respectively.

\begin{figure}[h!]
\centering
\includegraphics[width=0.99\columnwidth]{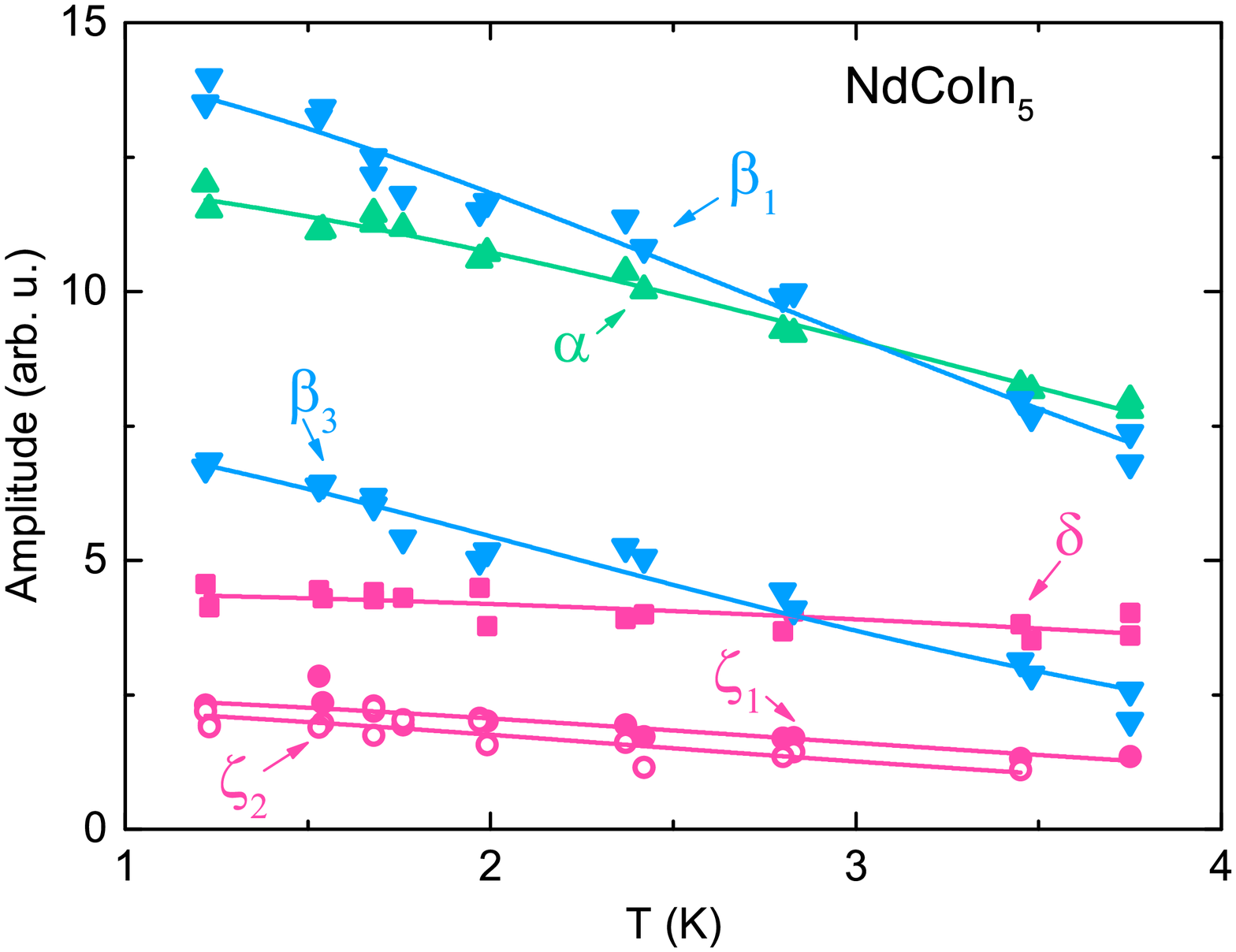}
\caption{Temperature-dependent dHvA oscillation amplitudes of NdCoIn$_5$ for both up and down sweeps between 21 and 34.5 T, taken at an angle $\theta=10^\circ$ off the $c$ axis. 
Lines are fits using the Lifshitz-Kosevich formula \cite{Shoenberg}.}
\label{fig:Fig5b}
\end{figure}

\section{Sommerfeld coefficients}

Estimated Sommerfeld coefficients from previously published specific-heat measurements are $\gamma\approx 0.36(1)$~J/(mol\,K$^2$) for Nd-2$\%$ and 5$\%$, as well as 0.37(1)~J/(mol\,K$^2$) for Nd-10$\%$ 
\cite{Hu08, note_2}, which is surprisingly large given the comparably small effective masses observed in our dHvA study (Tables I and II in the main text). 
It could be speculated that the reason for these large Sommerfeld coefficients for Nd doping levels 10$\%$ and less, but light effective masses is there are ``heavy'' orbits present, such as the $\beta$ orbits in CeCoIn$_5$ \cite{Settai01, Hall01}, that could not be seen under the used experimental conditions and would require the use of much larger magnetic fields to be observed experimentally. 
Another explanation is the existence of local quasiparticles that contribute to the heat capacity, but not to the FS.
These non-coherent quasiparticles would most likely be Kondo singlets and are commonly seen in other heavy-fermion compounds \cite{Zwicknagl16, Zwicknagl05}. 
A third explanation is the Sommerfeld coefficients were extracted from specific-heat measurements at zero applied magnetic field, but the effective mass is known to be strongly dependent on magnetic field \cite{McCollam05}. 
It would be interesting to measure the specific heat above 18~T to see if the Sommerfeld coefficients are field dependent.

\end{document}